\documentclass[prl,twocolumn,showpacs,preprintnumbers,amsmath,amssymb,citeautoscript]{revtex4-1}
\usepackage{amssymb,amsfonts,amsmath}
\usepackage{graphicx}
\usepackage{bm}

\begin{document}

\title{Cyclotron Resonance in the Hidden-Order Phase of URu$_2$Si$_2$}

\author{S. Tonegawa$^{1}$, K. Hashimoto$^{1,*}$, K. Ikada$^{1}$, Y.-H. Lin$^{1}$, H. Shishido$^{1,\dag}$, Y. Haga$^{2}$, \\
T.\,D. Matsuda$^{2}$, E. Yamamoto$^{2}$, Y. Onuki$^{2,3}$, H. Ikeda$^{1}$, Y. Matsuda$^{1}$, and T. Shibauchi$^{1,\ddag}$}
\affiliation{
$^1$Department of Physics, Kyoto University, Kyoto 606-8502, Japan\\
$^2$Advanced Science Research Center, Japan Atomic Energy Agency, Tokai 319-1195, Japan\\
$^3$Graduate School of Science, Osaka University, Toyonaka, Osaka 560-0043, Japan
}
\date{\today}


\begin{abstract}
We report the first observation of cyclotron resonance in the hidden-order phase of ultra-clean URu$_2$Si$_2$ crystals, which allows the full determination of angle-dependent electron-mass structure of the main Fermi-surface sheets. We find an anomalous splitting of the sharpest resonance line under in-plane magnetic-field rotation. This is most naturally explained by the domain formation, which breaks the fourfold rotational symmetry of the underlying tetragonal lattice. The results reveal the emergence of an in-plane mass anisotropy with hot spots along the [110] direction, which can account for the anisotropic in-plane magnetic susceptibility reported recently. This is consistent with the `nematic' Fermi liquid state, in which itinerant electrons have unidirectional correlations. 
\end{abstract}

\maketitle


The nature of the hidden order (HO) in the heavy-electron metal URu$_2$Si$_2$ is a long-standing mystery \cite{Myd11} since the discovery of the phase transition at $T_{\rm HO}=17.5$\,K \cite{Pal85,Map86,Sch86}. 
There are several unique features that appear to be clues for understanding the HO phase. Below $T_{\rm HO}$, an electronic excitation gap is formed on a large portion of the Fermi surface (FS) and most of the carriers disappear \cite{Beh05,Kas07}.  The gap formation also occurs in the magnetic excitation spectra \cite{Wie07}. The HO ground state with no large static moment \cite{Bro87,Mat01,Tak07} changes to the large-moment antiferromagnetic state upon applying hydrostatic pressure \cite{Ami07}, but the resolved part of the FS \cite{Ohk99,Has10,Aok11,San09,Yos10,Sch10,Ayn10,Kaw11,Bia09} has a striking similarity between these different phases \cite{Has10,Aok11}, implying that the HO is nearly degenerate with the antiferromagnetic order. The magnetic torque measurements reveal the in-plane anisotropy of magnetic susceptibility \cite{Oka11}, suggesting some hidden mechanism which involves a twofold symmetry in the tetragonal $ab$ plane of the HO phase.

Despite the above peculiar signatures, however, the fundamental question ``what is the nature of the order parameter in the HO phase?'' remains open, mainly because a detailed knowledge of the FS topology in the HO is lacking. In fact, quantum oscillation experiments \cite{Ohk99,Has10,Aok11} have revealed the existence of small pockets in the HO phase similar to those in the antiferromagnetic phase, but the total density of states of these pockets is significantly smaller than the estimate from the electronic specific heat, indicating that there must be some missing FS sheets with heavy mass in the HO phase. 

Cyclotron resonance (CR) is a powerful probe of the detailed FS structure, which is a complementary technique to the quantum oscillation experiments. The CR stems from the transition between Landau levels formed by the quantized cyclotron motion of the conduction electrons.   It quantifies directly the effective mass $m^*_{\rm CR}$ of electrons moving along extremal orbits on FS sheets through the simple relation $m^*_{\rm CR}=eH_{\rm CR}/\omega$, where $\omega=2\pi f$ is the microwave angular frequency and $H_{\rm CR}$ is the resonance field. It should be noted that in one-component translationally invariant systems $m^*_{\rm CR}$ is not renormalized by the electron-electron interaction (the Kohn's theorem \cite{Koh61}), but in solids this theorem can be violated \cite{Kan97} especially for the heavy-fermion systems with interacting conduction and $f$ electrons \cite{Var86} and for multiband systems \cite{Yos05,Kim11}. Therefore the momentum dependence of $m^*_{\rm CR}$ in each FS should contain important information on the electron correlations, which sometimes are a source of emergent novel phases. Although there are several studies on CR in strongly correlated materials \cite{Kim11,Hil00,Rze02}, no observations of CR in heavy-fermion compounds have been reported so far. 

Here we report on the first observation of CR in URu$_2$Si$_2$, which provides the full determination of the main FS sheets including the missing heavy band. The in-plane angle dependence of $m^*_{\rm CR}$ shows an unexpected splitting for the sharpest resonance line, which does not match the topology of the FS structure calculated for a state with the fourfold symmetry. The results indicate the emergence of hot spots as well as domain formation which breaks the tetragonal symmetry. This offers a natural explanation for the in-plane anisotropy of magnetic susceptibility \cite{Oka11}. We infer that the fourfold rotational symmetry breaking in the electronic structure is a characteristic signature in the HO phase, providing a stringent constraint on the symmetry of the hidden order.

\begin{figure}[t]
\includegraphics[width=\linewidth]{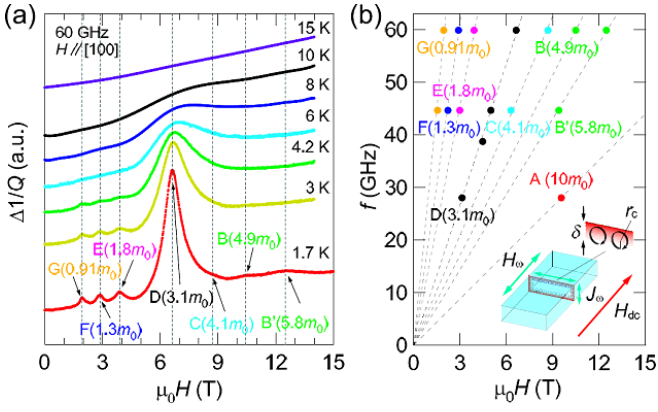}
\caption{(Color online) Observation of cyclotron resonance in the hidden-order phase of URu$_2$Si$_2$. (a) Field dependence of the change in $1/Q$ at several temperatures for $\bm{H}\parallel [100]$. Each resonance field corresponds to the cyclotron mass of each extremal orbit as labeled in unit of the free electron mass $m_0$. Each curve is shifted vertically for clarity. (b) Resonance fields at different frequencies for $\bm{H}\parallel [100]$ show the linear relation (dotted lines). Inset shows a schematic configuration of the microwave measurements. The dc field $\bm{H}$ is applied parallel to the microwave field component $\bm{H}_\omega$, which excites the microwave current $\bm{J}_\omega$ near the sample surface in the region characterized by the skin depth $\delta$ (red shades). }
\label{Fig:CR}
\end{figure}

The CR experiments were carried out on a URu$_2$Si$_2$ single crystal (with dimensions of $2.1\times0.58\times0.10$\,mm$^3$) having very large residual-resistivity-ratio ($\sim700$ \cite{Mat11}), with dc magnetic field $\bm{H}$ perpendicular to the alternating currents $\bm{J}_\omega$ induced by microwave [inset of Fig.\:\ref{Fig:CR}(b)]. In a classical picture, the electrons undergo cyclotron motion with velocity perpendicular to $\bm{H}$ and are accelerated by the microwave. When the frequency of the cyclotron motion coincides with the microwave frequency, the CR occurs near the surface within the microwave skin depth $\delta=(2\rho/\mu_0\omega)^{1/2}$ ($\sim 0.3\,\mu$m for 28\,GHz at 1.7\,K where the dc resistivity $\rho$ is $\sim 1\,\mu\Omega$cm). We note that unlike conventional CR in metals \cite{Azb58}, the small carriers in URu$_2$Si$_2$ make the cyclotron radius $r_c=v_F/\omega\approx \hbar k_F/eH_{\rm CR}$ shorter than $\delta$ in our measurement frequency range, where $v_F$ is the Fermi velocity and $k_F$ is the Fermi momentum. We used the cavity perturbation technique \cite{Shibauchi97} to measure the change in the $Q$-factor of the cavity resonator due to the microwave absorption of the crystal. 


As shown in Fig.\:\ref{Fig:CR}(a), the microwave power dissipation $\Delta1/Q$ as a function of applied field exhibits multiple resonance peaks, and the resonance lines become broader with increasing temperature. Measurements at different microwave frequencies demonstrate that the resonance field is proportional to the frequency [Fig.\:\ref{Fig:CR}(b)]. These are clear signatures of the CR. The observed seven CR lines are labeled as A to G in the order of corresponding $m^*_{\rm CR}$ from the heaviest. The normalized full width at half maximum (FWHM) $\Delta H_{\rm CR}/H_{\rm CR}$ gives an estimate for $\omega_c\tau$ (where $\omega_c$ is the cyclotron angular frequency and $\tau$ is the scattering time), which reaches $\sim 20$ at low temperatures for the sharpest line D. 

\begin{figure}[t]
\includegraphics[width=\linewidth]{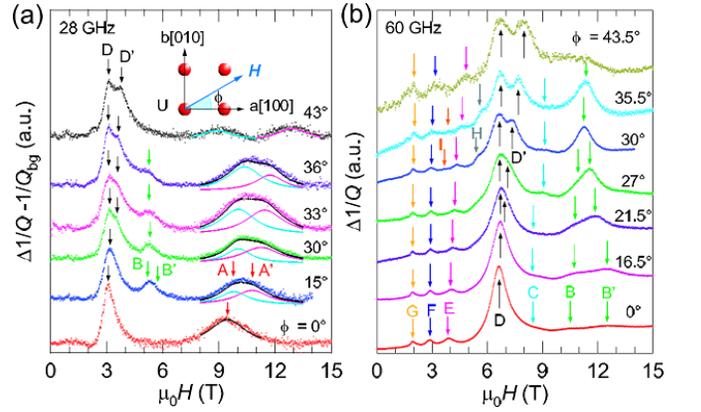}
\caption{(Color online) Cyclotron resonance under in-plane field rotation. (a) Microwave dissipation as a function of field measured at 28\,GHz. A smooth polynomial background field dependence $1/Q_{\rm bg}$ has been subtracted and the two-peak fitting analysis is used (lines) to extract the resonance fields A and A'. Inset shows schematic U atom arrangements in the tetragonal crystal structure and the definition of magnetic-field angle $\phi$ in the $ab$ plane. (b) The same plot for 60\,GHz. The arrows indicate the resonance fields. Each curve is shifted vertically for clarity. }
\label{Fig:in-plane}
\end{figure}

The field-angle dependence of the CR lines allows the determination of the angle-dependent electron masses on the FS sheets in the HO phase. Figures\:\ref{Fig:in-plane}(a) and \ref{Fig:in-plane}(b) display the resonance lines at two different frequencies when the field is inclined from $[100]$ toward $[110]$ direction in the $ab$ plane.  Among the observed CR lines, three lines A (A'), B, and D (D') have strong intensities, which should come from the main FS sheets with large volumes.  Resonance fields of the lines B and D at these two frequencies provide quantitatively consistent masses, indicating that the mass is field-independent at any angle within the measurement range of field. 

\begin{figure*}[t]
\includegraphics[width=0.85\linewidth]{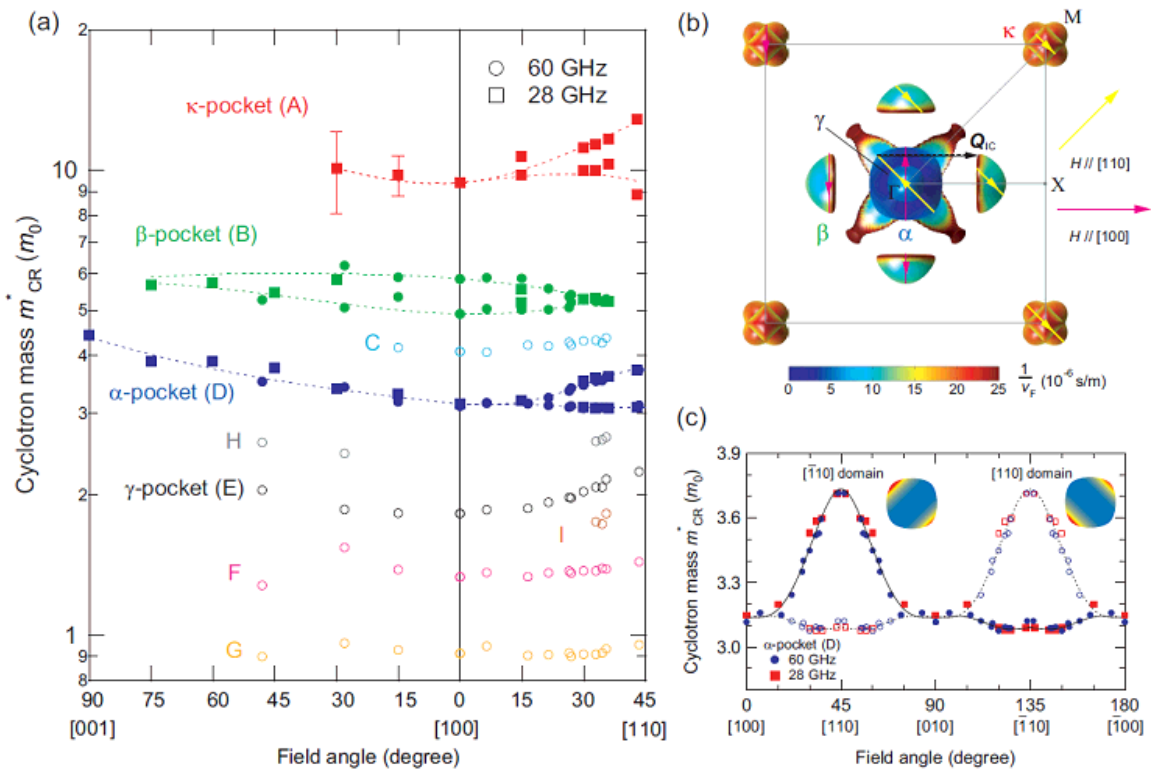}
\caption{(Color online) Structure of the cyclotron masses $m^*_{\rm CR}$ for each Fermi surface sheet in URu$_2$Si$_2$. (a) Cyclotron masses as a function of field angle for the $ac$-plane (left) and $ab$-plane (right) rotations. The solid (open) symbols are for the resonance lines with large (small) intensities. The dashed lines are guides to the eyes for the main three bands. (b) Schematic cross sectional view of FS in a plane including $\Gamma$ (the center), $X$, and $M$ points, obtained by the density functional band-structure calculations assuming the antiferromagnetic order \cite{Ike12}. The color indicates the inverse of Fermi velocity $1/v_F$ on the FS sheets. The extremal orbits relevant to the cyclotron resonance are indicated for the main $\alpha$, $\beta$ and $\kappa$ bands by yellow (pink) arrows for $\bm{H}\parallel [110]$ ($[100]$). The dashed arrow depicts the incommensurate wave vector $\bm{Q}_{\rm IC}$. 
(c) Cyclotron mass $m^*_{\rm CR}$ of the $\alpha$ band as a function of the in-plane field angle $\phi$ expected for each domain (solid (open) symbols for the $[\bar{1}10]$ ($[110]$) domain). The lines are the guides to the eyes. Insets show schematic mass distribution revealed by the CR data for the quasi-spherical $\alpha$ pocket viewed along the $c$ axis for the two domains. The red color represents the hot (heavy) spots.}
\label{Fig:mass}
\end{figure*} 

The angular dependence of the effective cyclotron masses obtained from the present CR measurements is summarized in Fig.\:\ref{Fig:mass}(a). Based on the fact that the FS in the HO phase is similar to that in the antiferromagnetic state \cite{Has10}, we compare our CR results with the band structure calculations assuming the antiferromagnetism \cite{Opp10,Ike12}. It has been shown by the density functional theory \cite{Ike12} that in the antiferromagnetic state, there are three non-equivalent FS sheets with relatively large volumes [Fig.\:\ref{Fig:mass}(b)] as well as several smaller pockets and cages that are likely responsible for the weak resonance lines [open symbols in Fig.\:\ref{Fig:mass}(a)].  The large main three pockets, labeled as $\alpha$ (hole), $\beta$ (electron), and $\kappa$ (electron), have distinctly different shapes and masses [Fig.\:\ref{Fig:mass}(b)]. We show that these three pockets are responsible for the observed three strong CR lines, D, B, and A, respectively [closed symbols in Fig.\:\ref{Fig:mass}(a)]. 

The $\alpha$ hole pocket with nearly isotropic shape, which locates around the center of the folded Brillouin Zone ($\Gamma$ point), has the largest volume. The positive Hall coefficient in this compensated system indicates that this hole pocket has much larger mobility than the other electron pockets $\beta$ and $\kappa$ \cite{Kas07}. This $\alpha$ pocket is therefore responsible for the line D having the strongest intensity and sharpest FWHM (with the largest $\omega_c\tau$). There are four $\beta$ electron pockets with hemispherical shape in the direction towards the $X$ point. These pockets yield two different extremal orbits for $\bm{H}\parallel [100]$ [pink vertical arrows in Fig.\:\ref{Fig:mass}(b)], but these two become equivalent for $\bm{H}\parallel [110]$ [yellow diagonal arrows in Fig.\:\ref{Fig:mass}(b)]. This uniquely corresponds to the angle dependence of the line B. This line B also tends to merge towards $\bm{H}\parallel [001]$, which further supports this assignment. The $\kappa$ sheets around $M$ point have much heavier band mass with larger $1/v_F$ than the $\alpha$ pocket [Fig.\:\ref{Fig:mass}(b)], which naturally leads us to assign the heaviest line A to the $\kappa$ sheets. Indeed, the $\kappa$ FS consists of two crossing sheets [Fig.\:\ref{Fig:mass}(b)], which should give two different orbits for in-plane fields except when the field is aligned exactly parallel to the $[100]$ direction as observed for line A.

Our assignments for the $\alpha$ and $\beta$ bands are consistent with the quantum oscillation reports \cite{Ohk99,Has10}, in which the largest amplitude oscillation branch is assigned to $\alpha$ \cite{Ohk99} and the merging branches for $[100]\rightarrow [001]$ to $\beta$ \cite{Has10}. We note that different band assignments to the quantum oscillation branches have been proposed \cite{Opp10}, in which the largest $\Gamma$-centered hole band is assigned to the $\epsilon$ branch observed only at very high fields above $\sim 17$\,T \cite{Shi09}. However, such a high-field branch is most likely associated with field-induced transition \cite{Shi09} possibly due to the Lifshitz topology change by the Zeeman effect \cite{Jo07,Alt11}. It is rather reasonable to assign the most pronounced $\alpha$ branch to this largest hole band, which is consistent with our assignments of CR lines. We also stress that our sharpest line D cannot be assigned to $\kappa$, because no splitting is observed for the $ac$ rotation toward $[001]$. Moreover, from our assignments of the main bands (together with the small electron pocket $\gamma$ assigned to the CR line E) we can estimate the total electronic specific heat coefficient to be $\sim 53$\,mJ/K$^2$mol \cite{gamma} which accounts for more than 80\% of the experimental value of $\sim 65$\,mJ/K$^2$mol \cite{Map86}. Based on these results we conclude that the main FS sheets are now fully determined including the heaviest electron band $\kappa$ that has been missing in the quantum oscillation measurements, which detected only about half of the enhanced mass \cite{Has10}. The heaviest mass and small mean free path of the $\kappa$ pocket as revealed by the large width of line A are likely responsible for the difficulty in observing the corresponding oscillation frequency. 


Having established the assignments between the main CR lines and main FS pockets, we now focus on the signature of the FS that provides a key to understanding the HO.  The most unexpected and important result is that the sharpest line D arising from the $\alpha$ hole pocket is clearly split into two lines with nearly equal intensities near the $[110]$ direction [Figs.\:\ref{Fig:in-plane}(a), \ref{Fig:in-plane}(b) and \ref{Fig:mass}(a)]. This splitting in the $\alpha$ hole pocket is hard to explain from the calculated FS structure in the antiferromagnetic phase. One may argue that some small warping of FS shape gives rise to the appearance of an additional extremal orbit for particular angles, which can result in the splitting. However, such a scenario is highly unlikely because of the following reasons.  First, quantum oscillation measurements clearly indicate that the number of extremal orbits in the $[110]$ direction remains the same as that in $[100]$ direction \cite{Ohk99,Aok11}. We should add that in these measurements for $\bm{H}\perp[001]$ multiple frequencies with a constant separation are observed, which likely has a magnetic breakdown origin and is totally different from the splitting behavior observed in the present CR measurements.   Second, the band-structure calculation indicates that $\alpha$ pocket has nearly isotropic shape, which is supported by the quantum oscillations experiments \cite{Ohk99,Has10} that reveal almost angle-independent oscillation frequency for this band.  Third, the fact that the integrated intensity of the split line D' is nearly equal to that of D line in a wide range of angle [Figs.\:\ref{Fig:in-plane}(a) and (b)] suggests that both CR lines arise from the orbits with nearly equal FS cross sections, which is at odds with the warping scenario. Therefore we infer that the observed splitting of the CR line D results from a peculiar mass structure in the HO phase. 

A cyclotron mass is given by the inverse Fermi velocity integrated over an extremal orbit as $\frac{\hbar}{2\pi}\oint\frac{\mathrm{d}\bm{k}}{v_F}$, and thus we need two different orbits to have a mass splitting. The observed splitting is most naturally accounted for by the in-plane mass anisotropy which has heavy (hot) spots only near the orbit for $\bm{H}\parallel [110]$ [insets in Fig.\:\ref{Fig:mass}(c)]. This breaks the fourfold tetragonal symmetry of the lattice, which leads to the two domains inside the crystal, one along $[110]$ and the other along $[\bar{1}10]$ direction owing to their degenerate nature. These two domains give rise to two different cyclotron masses as observed in the present measurements [Fig.\:\ref{Fig:mass}(c)]. The comparable intensities of the split peaks D and D' imply similar volumes of the two domains, which is reasonable for the macroscopic crystal used in this study.  

Our CR results provide evidence that the HO phase is a `nematic' Fermi liquid of strongly correlated itinerant electrons, in which the electronic structure breaks the rotational fourfold symmetry of the underlying crystal lattice \cite{Fra10}. The quasi-spherical hole band has unidirectional hot spots with a $\sim 20$\% mass enhancement along the $[110]$ ($[\bar{1}10]$) direction in each domain [Fig.\:\ref{Fig:mass}(c)]. This indicates that the slope of the energy-momentum dispersion near the $\alpha$ FS can be different between $[110]$ and $[\bar{1}10]$ directions due to correlation effect, while the size of Fermi momentum remains nearly the same. The broken fourfold symmetry in the FS consistently accounts for the magnetic susceptibility difference between $[110]$ and $[\bar{1}10]$ directions reported by the torque measurements in tiny single crystals \cite{Oka11}. We stress that the torque signals of two domains cancel each other, but the CR in a large crystal can detect both contributions, resulting in bulk evidence that itinerant electrons have nematic correlations. 

The deduced electronic mass structure imposes stringent constraints on the theories of the HO \cite{Myd11,Ike12,Kis05,Var06,Hau09,Cri09,Har10,Tha11,Pep11,Fuj11,Opp11,Cha12,Rau12}. Among the 10 symmetries allowed for URu$_2$Si$_2$ \cite{Kis05}, the broken fourfold symmetry of the electronic structure immediately restricts the symmetries of the order parameter to the doubly degenerate $E$ representation \cite{Tha11}, which has two symmetries $E^{+}$ and $E^{-}$ exhibiting the even ($+$) and odd ($-$) parities with respect to time reversal. Indeed, recent theories proposed order parameters with $E^{-}$ symmetry \cite{Ike12,Cha12,Rau12}. Another important implication is that the presence of the unidirectional hot spots on the FS implies that the electron correlations have strong momentum dependence. The hot spots in the $\alpha$ hole band can be connected to the edges of the $\beta$ electron pockets approximately by the incommensurate wave vector $\bm{Q}_{\rm IC}=(0.4, 0, 0)$ [dashed arrow in Fig.\:\ref{Fig:mass}(b)], whose excitations are found to be important by neutron scattering \cite{Wie07,Bro87}. The rotational symmetry breaking may enhance the interband scattering at the two of the four corners of the $\alpha$ band, generating the unidirectional hot spots as found in the present CR measurements. 

In summary, we have presented our CR data in the HO phase of URu$_2$Si$_2$, from which we fully determine the FS mass structure of the main bands. We found a clear splitting of $m^*_{\rm CR}$ for the $\alpha$ hole band, which is unexpected from the band-structure calculations assuming the fourfold symmetry. The results are consistent with a nematic Fermi liquid with the unidirectional Fermi-surface hot spots along the $[110]$ direction.

We thank D. Aoki, K. Behnia, 
P. Chandra, P. Coleman, S. Fujimoto, 
G. Knebel, H. Kontani, G. Kotliar, Y. Kuramoto, 
J.\,A. Mydosh, K. Miyake, 
M.-T. Suzuki, T. Takimoto, P. Thalmeier, C.\,M. Varma, 
and H. Yamagami for helpful discussions. This work was supported by Grant-in-Aid for the Global COE program ``The Next Generation of Physics, Spun from  Universality and Emergence'', Grant-in-Aid for Scientific Research on Innovative Areas ``Heavy Electrons'' (No.\,20102002, 20102006, 23102713) from MEXT, and KAKENHI from JSPS. 


\begin{thebibliography}{999}
\bibitem{gamma}
For the Sommerfeld coeficient estimate, we use the quantum oscillations results \cite{Ohk99,Has10} for the volumes and masses (for $\bm{H}\parallel [001]$) of $\alpha$, $\beta$, and $\gamma$ pockets. The CR mass $m^*_{\rm CR}$ in general may be different from the quantum oscillation mass $m^*_{\rm QO}$ \cite{Kan97}, and we find the ratio $m^*_{\rm QO}/m^*_{\rm CR}$ is in a range of $3-4$ for these pockets, and thus we use the ratio of 4 to estimate the thermodynamic mass of the $\kappa$ pockets. The volume of $\kappa$ is estimated from the compensation condition.
\end{thebibliography}
\end{document}